\documentclass[twocolumn,twoside,slac]{revtex4}
\usepackage{graphicx}
\usepackage{fancyhdr}
\usepackage{epsfig}
\usepackage{axodraw}
\usepackage{amsfonts}
\usepackage{amssymb}
\begin{document}

%Title of paper
\title{Frequentist Hypothesis Testing with Background Uncertainty}

% Repeat the \author .. \affiliation  etc. as needed
%
% \affiliation command applies to all authors since the last
% \affiliation command. The \affiliation command should follow the
% other information

\author{K.S. Cranmer}
\affiliation{University of Wisconsin-Madison, Madison, WI 53706, USA}

\begin{abstract}
We consider the standard Neyman-Pearson hypothesis test of a
signal-plus-background hypothesis and background-only hypothesis in
the presence of uncertainty on the background-only prediction.
Surprisingly, this problem has not been addressed in the recent
conferences on statistical techniques in high-energy physics --
although the its confidence-interval equivalent has been.  We discuss
the issues of power, similar tests, coverage, and ordering rules.  The
method presented is compared to the Cousins-Highland technique, the
ratio of Poisson means, and ``profile'' method.
\end{abstract}

%\maketitle must follow title, authors, abstract
\maketitle

\thispagestyle{fancy}

\section{Introduction\label{S:intro}}

In the last five years there have been several conferences on
statistics for particle physics.  Much of the emphasis of these
conferences were on limit setting and the Feldman-Cousins ``unified
approach'', the quintessential frequentist method based on the Neyman
construction.  As particle physicists prepare for the Large Hadron
Collider (LHC) at CERN, we will need to reexamine our list of
statistical tools in the context of discovery.  In fact, there has
been no presentation at these statistical conferences on frequentist
hypothesis testing in the presence of uncertainty on the background.
%Furthermore, the search for the Higgs boson at LEP was based on the
%Cousins-Highland technique for incorporating systematic errors, which
%has a fundamentally Bayesian component.

In Section~\ref{S:HypothesisTesting} we will review the Neyman-Pearson
theory for testing between two simple hypotheses, and examine the
impact of background uncertainty in
Section~\ref{S:NuisanceParameters}.  In
Sections~\ref{S:NeymanConstruction}-~\ref{S:OrderingRule} we will
present a fully frequentist method for hypothesis testing with
background uncertainty based on the Neyman Construction.  In the
remainder of the text we will present an example and compare this
method to other existing methods.

\section{Simple Hypothesis Testing\label{S:HypothesisTesting}}

In the case of Simple Hypothesis testing, the Neyman-Pearson theory
(which we review briefly for completeness) begins with two Hypotheses:
the null hypothesis $H_0$ and the alternate hypothesis
$H_1$~\cite{Kendall}.  These hypotheses are called {\it simple}
because they have no free parameters.  Predictions of some physical
observable $x$ can be made with these hypotheses and described by the
likelihood functions $L(x|H_0)$ and $L(x|H_1)$ (for simplicity, think of $x$ as
the number of events observed).

Next, one defines a region $W\in I$ such that if the data fall in $W$
we accept the $H_0$ (and reject $H_1$).  Conversely, if the data fall
in $I -W$ we reject $H_0$ and accept the $H_1$.  The probability to
commit a Type I error is called the {\it size} of the test and is
given by
\begin{equation}\label{E:typeI}
\alpha = \int_{I-W} L(x|H_0) dx.
\end{equation}
The probability to commit a Type II error is given by
\begin{equation}
\beta = \int_W L(x|H_1) dx.
\end{equation}
Finally, the Neyman-Pearson lemma tells us that the region $W$ of size
$\alpha$ which minimizes the rate of Type~II error (maximizes the
power) is given by
\begin{equation}
W = \left \{x ~ \Bigg | ~\frac{L(x|H_1)}{L(x|H_0)} > k_\alpha \right \}.
\end{equation}

\section{Nuisance Parameters\label{S:NuisanceParameters}}
%\section{Similar Tests, Coverage, and Power}

Within physics, the majority of the emphasis on statistics has been on
limit setting -- which can be translated to hypothesis testing through
a well known dictionary~\cite{Kendall}.  When one includes nuisance
parameters $\theta_s$ (parameters that are not of interest or not
observable to the experimenter) into the calculation of a confidence
interval, one must ensure coverage for every value of the nuisance
parameter.  When one is interested in hypothesis testing, there is no
longer a physics parameter $\theta_r$ to cover, instead one must
ensure the rate of Type I error is bounded by some predefined value.
Analogously, when one includes a nuisance parameters in the null
hypothesis, one must ensure that the rate of Type I error is bounded
for every value of the nuisance parameter.  Ideally one can find an
acceptance region $W$ which has the same size for all values of the
nuisance parameter ({\it i.e.} a similar test).  Furthermore, the
power of a region $W$ also depends on the nuisance parameter; ideally,
we would like to maximize the power for all values of the nuisance
parameter ({\it i.e.} Uniformly Most Powerful).  Such tests do not
exist in general.

In this note, we wish to address how the standard hypothesis test
is modified by uncertainty on the background prediction.  The
uncertainty in the background prediction represents the presence of a
nuisance parameter: for example, let us assume it is the expected
background $b$.  Typically, an auxiliary, or side-band, measurement is
made to provide a handle on the nuisance parameter.  Let us
generically call that measurement $M$ and $L(M|H_0,b)$ the prediction
of that measurement given the null hypothesis with nuisance parameter
$b$.  In Section~\ref{S:RatioPoisson} we address the special case that
$L(M|H_0,b)$ is a Poisson distribution.

\section{The Neyman-Construction\label{S:NeymanConstruction}}

Usually one does not consider an explicit Neyman construction when
performing hypothesis testing between two simple hypotheses; though
one exists implicitly.  Because of the presence of the nuisance
parameter, the implicit Neyman construction must be made explicit and
the dimensionality increased.  The basic idea is that for each value
of the nuisance parameters $\theta_s$, one must construct an
acceptance interval (for $H_0$) in a space which includes their
corresponding auxiliary measurements $M$, and the original test
statistic $x$ which was being used to test $H_0$ against $H_1$.

For the simple case introduced in the previous section, this requires
a three-dimensional construction with $b$, $M$, and $x$.  For each
value of $b$, one must construct a two-dimensional acceptance region
$W_b$ of size $\alpha$ (under $H_0$).  If an experiment's data $(x_0,
M_0)$ fall into an acceptance region $W_b$, then one cannot exclude
the null hypothesis with $100(1-\alpha)\%$ confidence.  Conversely, to
reject the null hypothesis ({\it i.e.} claim a discovery) the data
must not lie in any acceptance region $W_b$.  Said yet another way, to
claim a discovery, the confidence interval for the nuisance
parameter(s) must be empty (when the construction is made assuming the
null hypothesis).

\section{The Ordering Rule\label{S:OrderingRule}}

The basic criterion for discovery was discussed abstractly in the
previous section.  In order to provide an actual calculation, one must
provide an ordering rule: an algorithm which decides how to chose the
region $W_b$.  Recall, that there the constraint on Type~I error does
not uniquely specify an acceptance region for $H_0$.  In the
Neyman-Pearson lemma, it is the alternate hypothesis $H_1$ that breaks
the symmetry between possible acceptance regions.  Also in the unified
approach, it is the likelihood ratio that is used as an ordering
rule~\cite{Feldman:1998qc}.

At the Workshop on conference limits at FermiLab, Feldman showed that
Unified Method with Nuisance Parameters is in Kendall's Theory (the
chapter on likelihood ratio tests \& test
efficiency)~\cite{Feldman:fermilab}.  The notation used by Kendall is
given in Table~\ref{tab:Kendall}.  Also, Kendall identifies $H_0$ with
$\theta_r = \theta_{r0}$ and $H_1$ with $\theta_r \ne \theta_{r0}$.

\begin{table}
\begin{tabular}{ll}
Variable & Meaning \\ \hline
$\theta_r$ & physics parameters \\
$\theta_s$ & nuisance parameters\\
$\hat{\theta}_r, \hat{\theta}_s$ & unconditionally maximize $L(x|\hat{\theta}_r,\hat{\theta}_s)$\\
$\hat{\hat{\theta}}_s$ & conditionally maximize $L(x|\theta_{r0},\hat{\hat{\theta}}_s)$\\
\end{tabular}
\caption{The notation used by Kendall for likelihood tests with nuisance parameters}
\label{tab:Kendall}
\end{table}

Let us briefly quote from Kendall:
\begin{quote}
``Now consider the Likelihood Ratio
\begin{equation}\label{E:OrderingRule}
l = \frac{L(x|\theta_{r0},\hat{\hat{\theta}}_s)}{L(x|\hat{\theta}_r,\hat{\theta}_s)}
\end{equation}
Intuitively $l$ is a reasonable test statistic for $H_0$:
it is the maximum likelihood under $H_0$ as a fraction of its largest
possible value, and large values of $l$ signify that $H_0$ is
reasonably acceptable.''
\end{quote}

Feldman uses this chapter as motivation for the profile method (see
Section~\ref{S:Profile}), though in Kendall's book the same likelihood
ratio is used as an ordering rule {\it for each value of the nuisance
parameter}.  

The author tried simple variations on this ordering rule before
rediscovering it as written.  It is worth pointing out that
Eq.~\ref{E:OrderingRule} is independent of the nuisance parameter $b$;
however, the contour of $l_\alpha$ which provides an acceptance region
of size $\alpha$ is not necessarily independent of $b$.  It is also
worth pointing out that $\hat{\theta_r}$ and $\hat{\theta}_s$ do not
consider the null hypothesis -- if they did, the region in which $l=1$
may be larger than $(1-\alpha)$.  Finally, if one uses $\theta_s$
instead of $\hat{\theta}_s$ or $\hat{\hat{\theta}}_s$, one will not
obtain tests which are approximately similar.

\section{An Example}

Let us consider the case when the nuisance parameter is the expected
number of background events $b$ and $M$ is an auxiliary measurement of
$b$.  Furthermore, let us assume that we have a absolute prediction of
the number of signal events $s$.  For our test statistic we choose the
number of events observed $x$ which is Poisson distributed with mean
$\mu=b$ for $H_0$ and $\mu=s+b$ for $H_1$.  In the construction there
are no assumptions about $L(M|H_0,b)$ -- it could be some very
complicated shape relating particle identification efficiencies, Monte
Carlo extrapolation, etc.  In the case where $L(M|H_0,b)$ is a Poisson
distribution, other solutions exist (see Section~\ref{S:RatioPoisson}).
For our example, let us take $L(M|H_0,b)$ to be a Normal distribution
centered on $b$ with standard deviation $\Delta b$, where $\Delta$ is
some relative systematic error.  Additionally, let us assume that we
can factorize $L(x,M|H,b) = L(x|H,b)L(M|b)$ (where $H$ is either $H_0$
or $H_1$).

For our example problem, we can re-write the ordering rule in
Eq.~\ref{E:OrderingRule} as
\begin{equation}\label{E:OrderingRuleExample}
l = \frac{L(x,M|H_0,\hat{\hat{b}})}{L(x,M|H_1,\hat{b})},
\end{equation}
where $\hat{b}$ conditionally maximizes $L(x,M|H_1,b)$ and
 $\hat{\hat{b}}$ conditionally maximizes $L(x,M|H_0,b)$.

\begin{figure}
%\centerline{\epsfig{file=construction_fig.eps, width=.35\textwidth}}
\centerline{\epsfig{file=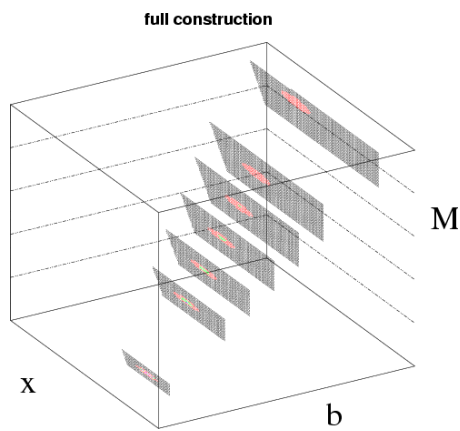, width=.35\textwidth}}
\caption{The Neyman construction for a test statistic $x$, an
auxiliary measurement $M$, and a nuisance parameter $b$.  Vertical
planes represent acceptance regions $W_b$ for $H_0$ given $b$.  The
condition for discovery corresponds to data $(x_0, M_0)$ that do not
intersect any acceptance region.  The contours of $L(x,M|H_0,b)$ are
in color.  }
\label{fig:construction}
\end{figure}

Now let us take $s=50$ and $\Delta = 5\%$, both of which could be
determined from Monte Carlo.  In our toy example, we collect data
$M_0=100$.  Let $\alpha= 2.85 \cdot 10^{-7}$, which corresponds to
$5\sigma$.  The question now is how many events $x$ must we observe to
claim a discovery?\footnote{In practice, one would measure $x_0$ and
$M_0$ and then ask, ``have we made a discovery?''.  For the sake of
explanation, we have broken this process into two pieces.}  The
condition for discovery is that $(x_0,M_0)$ do not lie in any
acceptance region $W_b$.  In Fig.~\ref{fig:construction} a sample of
acceptance regions are displayed.  One can imagine a horizontal plane
at $M_0=100$ slicing through the various acceptance regions.  The
condition for discovery is that $x_0>x_{\rm max}$ where $x_{\rm max}$
is the maximal $x$ in the intersection.
% of this horizontal plane and
%the acceptance regions $W_b$.

\begin{figure}
\centerline{\epsfig{file=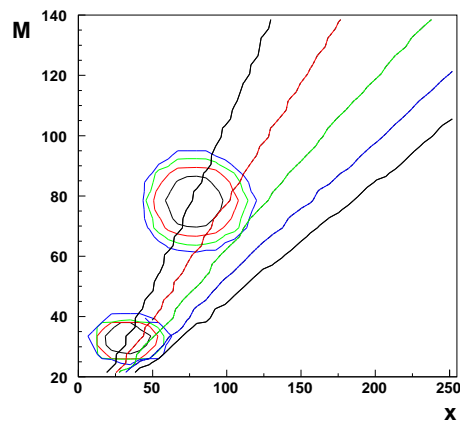, width=.35\textwidth}}
\caption{Contours of the likelihood $L(x,M|H_0,b)$ are shown as
concentric ellipses for $b=32$ and $b=80$.  Contours of the likelihood
ratio in Eq.~\ref{E:OrderingRuleExample} are shown as diagonal lines.  This
figure schematically illustrates that if one chooses acceptance
regions based solely on contours of the likelihood ratio, that
similarity is badly violated.  For example, data $M=80,x=130$ would be
considered part of the acceptance region for $b=32$, even though it
should clearly be ruled out.}\label{fig:contours}
\end{figure}

There is one subtlety which arises from the ordering rule in
Eq.~\ref{E:OrderingRuleExample}.  The acceptance region $W_b =
\{(x,M)~~|~~l>l_\alpha\}$ is bounded by a contour of the likelihood
ratio and must satisfy the constraint of size: $\int_{W_b}
L(x,M|H_0,b) = (1-\alpha)$.  While it is true that the likelihood is
independent of $b$, the constraint on size {\it is} dependent upon
$b$.  Similar tests are achieved when $l_\alpha$ is independent of
$b$.  The contours of the likelihood ratio are shown in
Fig.~\ref{fig:contours} together with contours of $L(x,M|H_0,b)$.
While tests are roughly similar for $b\approx M$, similarity is
violated for $M \ll b$.  This violation should be irrelevant because
clearly $b \ll M$ should not be accepted.  This problem can be avoided
by clipping the acceptance region around $M=b \pm N\Delta b$, where
$N$ is sufficiently large ($\approx 10$) to have negligible affect on
the size of the acceptance region.  Fig.~\ref{fig:construction} shows
the acceptance region with this slight modification.

In the case where $s=50$, $\Delta=5\%$, and $M_0 = 100$, one must
observe 167 events to claim a discovery.  While no figure is provided,
the range of $b$ consistent with $M_0=100$ (and no constraint on $x$)
is $b\in[68,200]$.  In this range, the tests are similar to a very high
degree.

%\section{Comparison with Other Methods}

%We now compare this method with the most common techniques.  These
%comparisons include practical, numerical, and philosophical
%differences.

\section{The Cousins-Highland Technique}

The Cousins-Highland approach to hypothesis testing is quite
popular~\cite{FinalLHWG:2003sz} because it is a simple smearing on the nuisance
parameter~\cite{Cousins:1992qz}.  In particular, the background-only
hypothesis $L(x|H_0,b)$ is transformed from a compound hypothesis with
nuisance parameter $b$ to a simple hypothesis $L'(x|H_0)$ by
\begin{equation}
L'(x|H_0) = \int_b L(x|H_0,b) L(b) db,
\end{equation}
where $L(b)$ is typically a normal distribution.  The problem with
this method is largely philosophical: $L(b)$ is meaningless in a
frequentist formalism.  In a Bayesian formalism one can obtain $L(b)$
by considering $L(M|b)$ and inverting it with the use of Bayes's
theorem and the {\it a priori} likelihood for $b$.  Typically,
$L(M|b)$ is normal and one assumes a flat prior on $b$.

In the case where $s=50$, $L(b)$ is a normal distribution with mean
$\mu = M_0 = 100$ and standard deviation $\sigma= \Delta M_0 = 5$, one
must observe 161 events to claim a discovery.  
%when using the Cousins-Highland method. 
Initially, one might think that 161 is quite
close to 167; however, they differ at the 4\% level and the methods
are only considering a $\Delta = 5\%$ effect.  Still worse, 
%is to
%realize that roughly 100 of these events come from the background
%process, and the additional number of events needed to claim a
%discovery (61 vs. 67) differ by approximately 10\%.  
if $H_0$ is true (say $b_t=100$) and one can claim a discovery with
the Cousins-Highland method ($x_0>161$), the chance that one could not
claim a discovery with the fully frequentist method $(x_0<167$) is
$\approx 95\%$.  Similarly, if $H_1$ is true %(with $b_t=100$)
 and one
can claim a discovery with the Cousins-Highland method, the chance
that one could not claim a discovery with the fully frequentist method
is $\approx 50\%$.  Even practically, there is quite a difference
between these two methods.

\section{The Ratio of Poisson Means\label{S:RatioPoisson}}

During the conference, J. Linnemann presented results on the ratio of
Poisson means%~\cite{Linnemann}
.  In that case, one considers a
background and a signal process, both with unknown means.  By making
``on-source'' ({\it i.e.} $x$) and ``off-source'' ({\it i.e.} $M$)
measurements one can form a confidence interval on the ratio $\lambda
= s/b$.  If the $100(1-\alpha)\%$ confidence interval for $\lambda$
does not include $0$, then one could claim discovery.  This approach
does take into account uncertainty on the background; however, it is
restricted to the case in which $L(M|b)$ is a Poisson distribution.

There are two variations on this technique.  The first technique has
been known for quite some time and was first brought to physics in
Ref.~\cite{James:1980}.  This approach conditions on $x+M$, which
allows one to tackle the problem with the use of a binomial
distribution.  Later, Cousins improved on these limits by removing the
conditioning and considering the full Neyman
construction~\cite{Cousins:1998}.  Cousins paper has an excellent
review of the literature for those interested in this technique.

\section{The Profile Method\label{S:Profile}}

As was mentioned in Section~\ref{S:NuisanceParameters} the likelihood ratio in
Eq.~\ref{E:OrderingRule} is independent of the nuisance parameters.
If it were not for the violations in similarity between tests, one
would only need to perform the construction for one value of the
nuisance parameters.  Clearly, $\hat{\hat{\theta}}_s$ is an
appropriate choice to perform the construction.  This is the logic
behind the profile method.  It should be pointed out that the profile
method is an approximation to the full Neyman construction; though a
particularly good one.  In the example above with $x_0 = 167$,
$M_0=100$, the construction would be made at $b=\hat{\hat{b}}=117$
which gives the identical result as the fully frequentist method.

The main advantage to the profile method is that of speed and
scalability.  Instead of performing the construction for every value
of the nuisance parameters, one must only perform the construction 
once.  For many variables, the fully frequentist method is not
scalable if one na{\"i}vely loops over on a fixed grid.  However,
Monte Carlo sampling the nuisance parameters does not suffer from the
curse of dimensionality and serves as a more robust approximation of
the full construction than the profile method.

\section{Conclusion}

We have presented a fully frequentist method for hypothesis testing.
The method consists of a Neyman construction in each of the nuisance
parameters, their corresponding auxiliary measurements, and the test
statistic that was originally used to test $H_0$ against $H_1$.  We
have chosen as an ordering rule the likelihood ratio with the nuisance
parameters conditionally maximized to their respective hypotheses.
With a slight modification, this ordering rule produces tests that are
approximately similar.  We have compared this method to the most
common methods in the field.  This method is philosophically more
sound than the Cousins-Highland technique and more general than the
ratio of Poisson means.  This method can be made computationally less
intensive either with Monte Carlo sampling of the nuisance parameters
or by the approximation known as the profile method.

% If you have acknowledgments, this puts in the proper section head.
\begin{acknowledgments}
This work was supported by a graduate research fellowship from the
National Science Foundation and US Department of Energy Grant
DE-FG0295-ER40896.  The author would like to thank L. Lyons,
R.D. Cousins, and G. Feldman for useful feedback.
\end{acknowledgments}

% Create the reference section using BibTeX:
%\bibliographystyle{apsrev}
\bibliographystyle{unsrt}
\bibliography{vbf,bruce_cites,stats,nn,PhysicsGP}

\end{document}